\documentclass[10pt,conference]{IEEEtran}
\IEEEoverridecommandlockouts
\usepackage[numbers,sort&compress]{natbib}

\usepackage{amsmath,amssymb,amsfonts}
\usepackage{graphicx}
\usepackage{textcomp}
\usepackage{xcolor}
\usepackage{booktabs}
\usepackage{array}
\usepackage{flushend}
\usepackage[hyphens]{url}
\usepackage[colorlinks,linkcolor=black,citecolor=black,urlcolor=black,bookmarks=false]{hyperref}
\begin{document}

\title{What Will Replace Pull Requests?\\Rethinking Software Collaboration for AI Agents}

\author{\IEEEauthorblockN{Christoph Treude}
\IEEEauthorblockA{Singapore Management University\\
Singapore\\
ctreude@smu.edu.sg}
\and
\IEEEauthorblockN{Sebastian Baltes}
\IEEEauthorblockA{Heidelberg University\\
Heidelberg, Germany\\
sebastian.baltes@uni-heidelberg.de}}

\maketitle

\begin{abstract}
The emergence of artificial intelligence (AI) coding agents has put pull requests under pressure: agents open millions of pull requests per month, while some open-source projects have begun to restrict or disable external contributions. We investigate which alternatives to pull requests practitioners are proposing or adopting and what these changes mean for empirical software engineering research. We reviewed 50 gray-literature documents from 31 organizations or individuals published between October 2025 and October 2026, coding the problems each named, the mechanisms it proposed or adopted, which of five roles of pull requests (integration, review, rationale, accountability, and external contribution) each mechanism takes over, and the evidence offered. Most responses retained pull requests in some form, whether by adding tooling, attaching a new artifact, or restricting access, and restriction was the response most often reported as adopted. Several proposed alternatives moved review and rationale into agent conversations and session records, but only Zed reported using a replacement in practice. Only two documents outlined a replacement for external contribution, and neither reported adoption. Pull requests are changing in ways that pull-request datasets will not capture: reviews conducted in agent conversations leave no pull-request comments, and projects that disable pull requests disappear from samples of newly opened ones. We outline a research agenda on the effectiveness of the emerging alternatives, their consequences for open-source participation, and the adaptation of empirical methods.
\end{abstract}


\section{Introduction}
\label{sec:intro}

\looseness=-1 Coding agents are generating changes at a scale that puts existing pull-request workflows under pressure. GitHub reported that agents were opening more than 17 million pull requests per month in June 2026~\cite{g60newstack}.
Faros AI, a developer analytics vendor, compared telemetry from 22,000 developers between each organization's periods of lowest and highest artificial intelligence (AI) adoption and reported a 441.5\% increase in median time in review and a 31.3\% increase in pull requests merged without any review~\cite{g19faros}.

\looseness=-1 The pressure is social as well as technical. Ghostty, a terminal emulator project, automatically closes pull requests from contributors whom a maintainer has not vouched for~\cite{g06ghostty}.
Ladybird, a web browser project, accepts code changes only from its maintainers~\cite{g56ladybird}.
The repository of OpenAI's Codex agent accepts no external code contributions because implementation ``is comparatively straightforward with the help of Codex itself''~\cite{g62codex}.
GitHub now lets maintainers cap the number of open pull requests from users without write access~\cite{g58prlimits} or turn pull requests off entirely~\cite{g14prsettings}.

\looseness=-1 In response, practitioners have started to discuss and build alternatives. Zed, the company behind the code editor of the same name, built Delta, a tool in which developers implement, review, and integrate changes within conversations with coding agents, without opening pull requests~\cite{g46deltabeta}.
Zed disabled pull requests on Delta's own repository, and by September 16, 2026, 33 people had committed 570 changes there~\cite{g46deltabeta}.
Entire keeps Git and saves the prompts and tool actions behind agent-generated code alongside commits~\cite{g11entire}.
We examine which problems these responses address, which roles of pull requests they take over, and which of them projects have adopted.

\looseness=-1 These developments also have consequences for empirical software engineering research, much of which mines pull requests. Pull requests give researchers a record of how contributions are proposed, reviewed, and accepted~\cite{gousios2014}.
\citeauthor{tsay2014} used that record to show that the amount of discussion on a pull request influenced its acceptance~\cite{tsay2014}.
Studies of coding agents rely on the same records: \citeauthor{li2026aidev} collected 932,791 agent-authored pull requests for the AIDev dataset~\cite{li2026aidev}.
When developers review a change inside an agent conversation, the pull request, if one exists at all, carries no record of that review. When a project turns pull requests off, it disappears from every sample of newly opened pull requests.

\looseness=-1 We put forward a research agenda for this shift, informed by a review of 50 gray-literature documents.

\section{Five roles of pull requests}
\label{sec:background}

\looseness=-1 GitHub's pull request combines the proposed branch and its diff with the discussion and a merge button~\cite{gousios2014}.
Contributors without commit rights fork the project and submit changes for maintainers to review and integrate~\cite{dabbish2012}.
\citeauthor{derosso2016}'s coherence criterion states that each concept of a system should be motivated by at most one purpose. They found that Git's commit violates it by both saving work and grouping logically related changes~\cite{derosso2016}.
We apply the same distinction to pull requests. Table~\ref{tab:roles} defines the five roles that we use to compare the reviewed responses, without claiming that they exhaust every function of pull requests. Most responses take over or restrict one role and leave the others to the pull request.

\begin{table}[t]
\caption{The five roles of pull requests used in this review, with what each role provides and an observation from prior work on pull-based development.}
\label{tab:roles}
\centering
\footnotesize
\setlength{\tabcolsep}{3pt}
\begin{tabular}{@{}>{\raggedright\arraybackslash}p{0.68in}>{\raggedright\arraybackslash}p{1.06in}>{\raggedright\arraybackslash}p{1.58in}@{}}
\toprule
Role & What the pull request provides & Observed in prior work \\
\midrule
Integration & Groups changes that are tested and merged together. & Integrators prioritized small, easy-to-integrate changes~\cite{gousios2015}. \\
Review & Lets reviewers inspect and discuss the changes. & Defect finding was the main motivation for review; knowledge transfer an added benefit~\cite{bacchelli2013}. \\
Rationale & Records why a change was made and why it was accepted or rejected. & Core members asked and contributors explained in comments~\cite{gousios2014}. \\
Accountability & Records who proposed and who approved a change; access rights govern merging. & Integrators trusted changes by the contributor's reputation~\cite{gousios2015}. \\
External contribution & Lets someone without write access propose a change. & Owners judged newcomers by visible activity and granted commit rights~\cite{dabbish2012}. \\
\bottomrule
\end{tabular}
\end{table}

\section{Method}
\label{sec:method}

\looseness=-1 We conducted a gray-literature review informed by \citeauthor{garousi2019}'s guidelines~\cite{garousi2019}.
The search began with coverage of Zed's Delta announcement in \emph{The New Stack}, a technology news publication~\cite{g47newstack}.
We followed links from each retrieved document until two consecutive documents added nothing new. On September 17, 2026, we also ran 32 web searches using terms such as ``pull request'' and ``GitHub alternative'' alongside terms related to agents and product names found during the search, and we repeated them on October 6 with eight follow-up queries. The second pass added 13 documents but no new problem or mechanism category. The supplementary material records the queries and screening decisions.

\looseness=-1 \emph{Selection.} We included documents published between October 2025 and October 6, 2026. The window opens with GitHub's announcement of Agent HQ~\cite{g02agenthq}.
Eligible documents either proposed or documented a change to contribution, review, or integration motivated by coding agents, or used primary data to document a failure of pull requests or GitHub under agent load. We excluded secondary coverage of an included source and documents whose original we could not retrieve. The searches yielded 63 candidates, of which 50 remained after coding. Twenty-two came from vendors of infrastructure, platforms, or review tools, nine were policies or statements by maintainers, ten were essays by practitioners, six came from press or analysts, and three were community discussion threads. Documents from the same organization or individual count as one source, for 31 sources in total. The first author also recorded each document's type (product announcement, vendor essay, practitioner essay, maintainer policy, trade press, industry report, or community discussion) to separate what has shipped or been adopted from what has been argued.

\looseness=-1 \emph{Coding.} A written codebook defined the labels (codes) and the rules for assigning them. We coded each document's stated problems using the five categories in Table~\ref{tab:problems}, its proposed or adopted mechanisms using six categories (Table~\ref{tab:unbundle}), the roles from Table~\ref{tab:roles} that the document assigned to each mechanism, its stance toward pull requests (keep, extend, replace, restrict, or bypass them, or leave GitHub), and the strongest evidence it offered. A mechanism was assigned only when the document described it in detail, and a role only when the document said that the mechanism takes over or supports it. For restriction, we recorded accountability when a document tied contribution to identity or vouching and external contribution when it closed or limited the channel. A document could be assigned several problems, and a mechanism several roles.

\looseness=-1 \emph{Checking.} Claude Fable 5.1 and GPT (gpt-6.1-sol) coded every candidate independently, each receiving the codebook and the full document texts with default settings (the supplementary material lists the prompts and batching). The two models agreed on 88\% of the 819 codes, with 99 disagreements across 49 documents. Cohen's $\kappa$ ranged from 0.46 for inclusion to 0.96 for mechanism. The first author, who had produced an initial AI-assisted coding as a comparison point, read the source documents to resolve every disagreement between the models (which excluded 13 of the 63 candidates) and every code on which both models agreed against the initial coding, and audited a random 15\% sample of the 550 codes on which all three agreed, changing one. The supplementary material contains the corpus, the model codes, each decision, and the agreement statistics for each dimension.

\looseness=-1 \emph{Limitations.} One author resolved the models' disagreements without a second human coder. Both models could make the same coding errors, so their agreement cannot establish that the classifications are correct. Claude's model family also helped draft the codebook and paper; GPT did not contribute to the codebook. Our search used one web search engine and began with links from a single article, so it may have favored English-language sources that link to one another. The corpus is not a representative sample of software projects: 22 of the 50 documents came from vendors, and the frequencies we report describe the reviewed documents, not adoption across the industry. We did not reproduce the vendors' measurements, and several products were in beta.

\begin{table}[t]
\caption{The five problems that the reviewed documents named, with the number of documents naming each problem and, in parentheses, the number of distinct sources. A document could name several problems.}
\label{tab:problems}
\centering
\footnotesize
\setlength{\tabcolsep}{3pt}
\begin{tabular}{@{}p{0.95in}p{1.85in}r@{}}
\toprule
Problem & Definition & Docs \\
\midrule
Review capacity & Reviewers cannot keep up with the changes. & 35 (24) \\
Lost provenance & The prompts, actions, and decisions behind agent-generated code are not preserved. & 22 (12) \\
Contribution norms & Expectations for contributions and the relationship between maintainers and contributors are unsettled. & 21 (17) \\
Accountability & It is unclear who is responsible for a change and under which policy. & 21 (17) \\
Volume and infrastructure & The platform or tooling cannot carry the amount of agent activity. & 15 (9) \\
\bottomrule
\end{tabular}
\end{table}

\section{Results}
\label{sec:results}

\subsection{Problems}

\looseness=-1 Table~\ref{tab:problems} lists the five problems. Review capacity was named most often in the reviewed documents and was the only problem named in every source category. Lost provenance was named mainly by vendors, including every document from Zed and Entire, whose products store agent sessions.
Contribution norms and accountability were the maintainers' problems. All nine maintainer documents named contribution norms, seven named accountability, and none named volume or infrastructure. The commit that introduced Ghostty's policy stated that ``agentic programming has eliminated the natural effort-based backpressure that previously limited low-effort contributions''~\cite{g06ghostty}.

\subsection{Responses}

\looseness=-1 In our corpus, 44 documents took a position on pull requests: nine would keep them and add tooling around them, nine would extend them with a new artifact, 11 would replace them with another unit of work, 12 would restrict who may submit them, two would leave GitHub while keeping pull requests, and one would dispense with pull requests and human review altogether. In 42 documents we found at least one of six mechanisms: \emph{threads}, \emph{prompt requests}, \emph{session checkpoints}, \emph{review prioritized by risk}, \emph{agent identity and scope}, and \emph{restriction}. Table~\ref{tab:unbundle} records which roles each mechanism took over. Stance and mechanism are separate codes: restriction was described in 13 documents, the 12 that would restrict access and one CodeRabbit essay that would keep pull requests~\cite{g39kravets}.

\begin{table}[t]
\caption{Number of documents in which each mechanism takes over, supports, or, for restriction, limits a role of pull requests: integration (Int.), review (Rev.), rationale (Rat.), accountability (Acc.), and external contribution (Ext.). A blank means none.}
\label{tab:unbundle}
\centering
\footnotesize
\setlength{\tabcolsep}{4.5pt}
\begin{tabular}{@{}lccccc@{}}
\toprule
Mechanism & Int. & Rev. & Rat. & Acc. & Ext. \\
\midrule
Thread & 4 & 7 & 8 & 1 & \\
Prompt request & & 6 & 3 & & 2 \\
Session checkpoint & & 7 & 7 & 2 & \\
Review prioritized by risk & & 14 & & & \\
Agent identity and scope & & & & 5 & \\
Restriction & & & & 10 & 13 \\
\bottomrule
\end{tabular}
\end{table}

\looseness=-1 \emph{Most responses retain pull requests.} In 20 of the 44 documents the pull request stayed the unit of work, with tooling around it, with a new artifact attached, or on a different host. The 12 that restricted access kept it as well. The tooling added around pull requests addresses review capacity and accountability. Review prioritized by risk appeared in 14 documents: Harness, a vendor of code hosting and software delivery tools, groups the changes within a pull request by risk rather than by file~\cite{g42harness}.
The two vendors leaving GitHub rebuilt hosting for agent traffic without changing the workflow: Cursor's Origin is ``designed for agent scale'' and ``Every repo has pull requests''~\cite{g37origin}, and Harness migrates ``pull requests, labels, webhooks, and branch rules'' from GitHub~\cite{g42harness}.
Rebuilding the infrastructure to carry more traffic is a different change from redesigning how people collaborate. Among the platform vendors, only Zed has shipped the latter, as a public beta~\cite{g46deltabeta}.

\looseness=-1 \emph{Several proposals move review and rationale into agent conversations and session records.} Threads, described in eight documents, and session checkpoints, described in nine, account for most of the review and rationale cells in Table~\ref{tab:unbundle}. In Zed's Delta, a reviewer opens a subthread with its own worktree, a working directory linked to the same Git repository, asks the agent to explain or revise a change, and leaves comments that remain attached to the code as it changes~\cite{g46deltabeta,g35delta}.
Entire keeps Git and, on each agent-generated commit, saves a session checkpoint holding the transcript, prompts, files touched, token usage, and tool calls on a separate branch~\cite{g11entire}.
A thinner record has already been adopted by two large projects. The Linux kernel asks that contributions made with AI tools carry an \texttt{Assisted-by} tag~\cite{g48kernel}. LLVM asks for the same trailer because labeling ``is intended to facilitate reviews''~\cite{g49llvm}.
Six documents went further and proposed reviewing the instructions given to the agent, four instead of the generated code and two before it. Propel, a review-tool vendor, proposed retaining diff review and adding structured information about the agent's goal, scope, constraints, tool permissions, and expected evidence~\cite{g23propel}.

\looseness=-1 \emph{Restriction is the response adopted most often in the reviewed documents.} Among the 12 documents that restricted pull requests, seven were policies that projects had adopted, four were GitHub's announcements and essays about its contribution controls, and one was a practitioner essay. Rust allows changes generated by language models only with disclosure and only when they are ``Pre-arranged, non-critical, high-quality, well-tested, and well-reviewed,'' and its reviewers may close non-compliant pull requests ``no questions asked''~\cite{g61rust}.
LLVM bans ``agents that take action in our digital spaces without human approval''~\cite{g49llvm}.
Ghostty, Ladybird, and the Codex repository close the channel to outsiders as described in Section~\ref{sec:intro}.
GitHub's cap on open pull requests from users without write access, with a bypass list for trusted contributors, turns the same policy into a platform setting~\cite{g58prlimits}.

\looseness=-1 \emph{External contribution remains unresolved.} Among the 11 documents that would replace pull requests, four came from Zed and one from press coverage of Zed, five were practitioner essays, and one was a maintainer's proposal. Only Zed reported use, by its own staff on Delta's repository. Zed itself still accepts pull requests for its editor; its founder explained: ``GitHub has an established social component that will take longer to replace, and we're not going to strand contributors to prove a point''~\cite{g47newstack}.
Two documents outlined how outsiders would contribute without pull requests, both through prompt requests; they are the two that assigned prompt requests the external-contribution role in Table~\ref{tab:unbundle}, whereas the other four treated the prompt as an input to review. The developer Pedro Pi{\~n}era suggested that project supporters could lend tokens to fund the execution of submitted prompts~\cite{g05pinera}.
Alejandro Revilla, founder of the payment software project jPOS, proposed that maintainers refine contributors' prompts and generate the implementation themselves under controlled conditions~\cite{g51jpos}.
Neither document reported a project using the workflow.

\subsection{Evidence}

\looseness=-1 Among the 44 documents that took a position, 15 offered only arguments, eight offered anecdotes, and six repeated numbers reported elsewhere. Most measurements came from vendors, such as CodeRabbit's comparison of issue counts in 470 open-source pull requests~\cite{g03coderabbit}.
Rust, one of two maintainers reporting its own project data, counted 1,281 open pull requests~\cite{g61rust}.
None of these measurements tested whether any mechanism in Table~\ref{tab:unbundle} improves review, and none of the 22 documents that named lost provenance tested whether session records help reviewers find defects.

\section{Implications}
\label{sec:implications}

\looseness=-1 \emph{For projects.} None of the reviewed mechanisms proposed as replacements for pull requests was documented as fulfilling all five roles in Table~\ref{tab:roles}. Restricting external contributions, the response adopted most often in the reviewed documents, may also reduce opportunities for the knowledge transfer that \citeauthor{bacchelli2013} found in code review~\cite{bacchelli2013}.
It may also reduce the visible activity through which project owners in \citeauthor{dabbish2012}'s study assessed newcomers and granted them commit rights~\cite{dabbish2012}.
Threads and session records serve review and rationale for people with write access and do not address outsiders.

\looseness=-1 \emph{What pull-request data measures.} Many studies decide whether a change was reviewed from the comments and approvals on its pull request. Comment datasets such as \citeauthor{sun2026}'s data on AI code review are collected the same way~\cite{sun2026}.
A change reviewed in an agent conversation or using a session checkpoint may have no pull-request comments documenting that review, causing studies that rely on those comments to misclassify it as unreviewed. Faros AI's 31.3\% rise in pull requests merged without review is computed from pull-request records~\cite{g19faros} and could appear to rise further as review moves elsewhere, even if every change were reviewed.

\looseness=-1 \emph{Who is in the sample.} Here the observations are missing rather than mismeasured. Projects that turn pull requests off, and changes integrated through a Delta thread without one, disappear from pull-request datasets. Projects that limit or cap pull requests remain represented, but with less external contribution. If projects that disable pull requests differ systematically from the rest, datasets of newly opened pull requests are biased. This extends \citeauthor{kalliamvakou2014}'s warning that GitHub does not capture all of a project's development activity~\cite{kalliamvakou2014}.

\looseness=-1 \emph{What data researchers would need.} The records that would restore these measures differ in how accessible they are. The \texttt{Assisted-by} trailers used by the Linux kernel and LLVM sit in commit messages~\cite{g48kernel}. Entire pushes session checkpoints to a branch inside the repository~\cite{g11entire}, whereas Delta threads are hosted by Zed, outside GitHub and its API~\cite{g46deltabeta}. GitHub's API reports a repository's current pull-request settings but not when they changed, so the date of a restriction has to be recovered from policy-file histories and announcements.
Where session records stay private, researchers will depend on partnerships with vendors or on studies that record the data themselves.

\section{Future Plans}
\label{sec:plans}

The review leaves three questions open.

\looseness=-1 \emph{How effective are the emerging alternatives for reviewing and integrating changes?} We will run an experiment in which developers review agent-generated changes under four conditions: the diff alone, the diff with the prompt that generated it, the diff with a short rationale summary, and the diff with the full session record. We will measure defects found, false positives, the correctness of acceptance decisions, and review time; the summary and full-record conditions test Propel's concern about asking reviewers to read long agent conversations~\cite{g23propel}.

\looseness=-1 \emph{What happens to open-source participation when contribution mechanisms change?} An event study of projects that restricted or disabled pull requests, compared with matched repositories that did not, would show how contribution and review activity change after a restriction and whether newcomers find other ways in. Interviews with maintainers on both sides of these policies and with external contributors would establish what prompted the policies and what maintainers need before accepting a change. We will also test whether existing contribution workflows without pull requests can accommodate agent-generated changes.

\looseness=-1 \emph{How should empirical software engineering methods adapt?} Researchers first need to know how much review already happens outside pull requests. Classic findings about pull requests, such as the effects of discussion and reputation on acceptance, should be re-examined in projects that moved review into threads. Mining studies also need a dated record of contribution-policy changes and measures of review that do not depend on pull-request comments.

\section{Conclusion}
\label{sec:conclusion}

\looseness=-1 What will replace pull requests? The question remains open. The alternatives we reviewed redistribute the roles that pull requests combine, but no implemented replacement in our corpus has demonstrated how external contributors would participate. The changes are nonetheless already under way: platforms have rebuilt their infrastructure for agent traffic, large projects have restricted who may contribute or introduced disclosure requirements for AI-generated code, and Zed has moved development of Delta out of pull requests. Software engineering researchers cannot wait for the answer, because the mechanism we study and the assumptions behind our pull-request datasets are being challenged at the same time.

\section*{Acknowledgments}
\looseness=-1 Generative AI disclosure: Anthropic's Claude (via Claude Code), including Claude Fable 5.1, assisted with drafting and revising all sections and the codebook. Claude Fable 5.1 and GPT (gpt-6.1-sol) coded the corpus, and a human author resolved every disagreement between them (Section~\ref{sec:method}). The authors reviewed and edited all generated material and take responsibility for the manuscript and artifacts.

\section*{Data availability}
\begingroup\raggedright The supplementary material is archived at \url{https://doi.org/10.5281/zenodo.23598329}.\par\endgroup

\bibliographystyle{IEEEtranN}

\end{document}